\documentclass[preprint2]{aastex}

\shorttitle{A Two-dimensional Map of Color Excess in NGC 3603}
\shortauthors{Pang et al.}

\begin{document}

\title{A Two-dimensional Map of the Color Excess in NGC 3603}


\author{Xiaoying Pang\altaffilmark{1}, Anna Pasquali, Eva K.\ Grebel}

\affil{Astronomisches Rechen-Institut, Zentrum f\"ur Astronomie der
          Universit\"at Heidelberg, M\"onchhofstr.\ 12--14, 69120
              Heidelberg, Germany}
 \email{xiaoying@ari.uni-heidelberg.de}


\altaffiltext{1}{Fellow of the International
          Max-Planck Research School for Astronomy and Cosmic
          Physics at the University of Heidelberg and of the
          Heidelberg Graduate School for Fundamental Physics}


\begin{abstract}

Using archival {\it HST}/WFC3 images centered on the young HD\,97950
star cluster in the giant H\,{\sc ii} region NGC\,3603, we computed
the pixel-to-pixel distribution of the color excess, $E(B-V)_{\rm g}$,
of the gas associated with this cluster from its H$\alpha$/Pa$\beta$
flux ratio. At the assumed distance of 6.9~kpc, the resulting median
color excess within 1~pc from the cluster center is $E(B-V)_{\rm g} =
1.51 \pm 0.04$ mag. Outside the cluster (at $r > 1$~pc), the color
excess is seen to increase with cluster-centric distance towards both
North and South, reaching a value of about 2.2~mag at $r = 2$~pc from
the cluster center. The radial dependence of $E(B-V)_{\rm g}$ westward
of the cluster appears rather flat at about 1.55~mag over the distance
range 1.2~pc $< r < 3$~pc.  In the eastern direction, $E(B-V)_{\rm g}$
steadily increases from 1.5~mag at $r = 1$~pc to 1.7~mag at $r =
2$~pc, and stays nearly constant at 1.7~mag for 2~pc $< r < 3$~pc. The
different radial profiles and the pixel-to-pixel variations of
$E(B-V)_{\rm g}$ clearly indicate the presence of significant
differential reddening across the 4.9~pc $\times 4.3$~pc area centered
on the HD\,97950 star cluster.  We interpret the variations of
$E(B-V)_{\rm g}$ as the result of stellar radiation and stellar winds
interacting with an inhomogeneous dusty local interstellar medium
(ISM) whose density varies spatially.  From the $E(B-V)_{\rm g}$
values measured along the rims of the prominent pillars MM1 and MM2 in
the southwest and southeast of the HD\,97950 cluster we estimate an
H$_2$ column density of $\log_{10}(N_{\rm H_2})=21.7$ and extrapolate
it to $\log_{10}(N_{\rm H_2})=23$ in the pillars' interior.  We find
the pillars to be closer to us than the central ionizing cluster and
suggest that star formation may be occurring in the pillar heads.  

\end{abstract}



\section{Introduction}

The young massive HD\,97950 star cluster in the giant H\,{\sc ii}
region NGC\,3603 is located in the Sagittarius-Carina arm of the Milky
Way.  This cluster is one of the most massive young star clusters in
the Galaxy ($\sim 10^4~M_\odot$, Harayama et al.\ 2008).  The cluster
hosts ten times more OB stars than the Orion Nebula Cluster (ONC),
including the two most massive binaries currently known in the Galaxy
(Schnurr et al.\ 2008).  The HD\,97950 cluster displays pronounced
mass segregation (e.g., Sung \& Bessell 2004; Harayama et al.\ 2008).
Earlier studies trying to age-date the HD\,97950 cluster found the
massive stars on the upper main sequence (MS) to be very young (e.g.,
1 -- 2~Myr according to the spectroscopic study by Melena et al.\
2008), while the pre-main-sequence stars (PMS) were found to show a
larger age spread of 2 -- 3 Myr (e.g., Eisenhauer et al.\ 1998; Grebel
2004, 2005; Harayama et al. 2008; Pang et al.\ 2010).  The PMS stars
in the outer cluster regions and in the surroundings of the cluster
may, in part, be older, but different authors arrive at different age
ranges (e.g., 4 -- 5 Myr; Sung \& Bessell 2004; Rochau et al.\ 2010)
or up to 10 Myr (Beccari et al.\ 2010).  Also the evolved supergiants
around the HD\,97950 cluster support the idea of a large age spread or
alternatively sequential or multiple episodes of star formation (e.g.,
Moffat et al.\ 1983; Tapia et al.\ 2001; Crowther et al.\ 2008; Melena
et al.\ 2008).  On the other hand, the size of the wind-blown bubble
around the cluster suggests an age much lower than 2.5 Myr when taking
the kinetic energy input of the massive stars into account (see
discussion in Drissen et al.\ 1995).  At present, the age(s) and the
star formation history of the  HD\,97950 star cluster and its
surroundings in NGC\,3603 are uncertain.  One of the uncertainties in
stellar age dating is introduced by the high and variable reddening in
NGC\,3603.  

The HD\,97950 star cluster lies in a wind-blown cavity north of a
large molecular cloud in the giant H\,{\sc ii} region NGC\,3603 (e.g.,
Clayton 1986; Melnick 1989).  The gaseous surroundings of the cluster
show a complex and variable velocity and density structure (e.g.,
Clayton 1990; Drissen et al.\ 1995).  Since the nebular density varies
spatially, one would expect that also the dust extinction changes with
position across the cluster area as denser clouds can shield dust from
stellar radiation better than less dense clouds. The resulting
variable reddening is known as differential reddening and was shown to
be present in NGC\,3603 (e.g., Sagar et al.\ 2001). When this effect
is not taken properly into account, it may make stars appear redder
and hence older in the color-magnitude diagram (CMD) depending on
their spatial position with respect to the line of sight.  This may
introduce a considerable uncertainty in the estimation of the stellar
ages.

Differential reddening could be one of the causes why the age spread
among PMS stars in the HD\,97950 star cluster appears to be as large
as up to 10 Myr. In fact, as shown by Beccari et al.\ (2010, their
Figure 7), PMS stars are spatially more widely distributed than the MS
stars and reside in areas with different reddening. Sung \& Bessel
(2004) find that the color excess of the stars within the HD\,97950
cluster core (at radii $r <$ 0.7 pc; Harayama et al.\ 2008) is
$E(B-V)_{\rm s}$ = 1.25 mag and rapidly increases to 2.1 mag in the
outer regions ($r \sim$ 12 pc).

We take advantage of the publicly available images of the HD\,97950
cluster taken with the Wide Field Camera 3 (WFC3) aboard the {\it
Hubble Space Telescope (HST)} through narrow-band filters centered on
the H$\alpha$ and Pa$\beta$ lines and through broad-band filters
sampling the continuum emission, in order to extend the work of Sung
\& Bessel (2004).  We compute a pixel-to-pixel map of the color
excess, $E(B-V)_{\rm g}$, across a $2.43' \times 2.14'$ ($4.9 \times
4.3$~pc) field centered on the HD\,97950 cluster. 
Since only some of the gas contributing to the reddening is located 
in front of the majority of the stars, the reddening of the gas that we derive is 
an upper limit to the true reddening experienced by the stars. 
 Thus the two-dimensional distribution of the color excess associated with the gas 
is the first step towards correcting individual MS and PMS 
stars for reddening in the WFC3 field of view. Multi-band stellar photometry
 will be needed in order to establish the conversion between 
the reddening of the gas and that of the stars. 
This will ultimately allow us to constrain more tightly the age spread
of PMS stars and the recent star formation history of the HD\,97950
star cluster and of its immediate surroundings (Pang et al.\ 2011, in
preparation). Our present paper focuses on the reddening map.  The
data and their reduction are described in Section 2, while the
two-dimensional map of the color excess is presented and discussed in
Section 3. Conclusions and a summary are presented in Section 4.

\section{Observations and Data Reduction}

Multi-wavelength imaging of the HD\,97950 star cluster was carried out
in 2010 with {\it HST}/WFC3 (proposal ID: 11360, PI: Robert
O'Connell). For our analysis we use the optical images taken in the
{\em F555W} ($\sim$ {\em V}), {\em F656N} ({\em H}$\alpha$), and {\em
F814W} ($\sim$ {\em I}) filters with the WFC3 UVIS detector (0.04
arcsec~pixel$^{-1}$) and near-infrared images in the {\em F127M}
(continuum in {\em J}), {\em F128N} ({\em Pa}$\beta$), and {\em F139M}
(continuum in {\em J}) filters obtained with the WFC3 IR detector
(0.13 arcsec~pixel$^{-1}$). The total exposure time in the different
filters is summarized in Table 1.  

All images were reduced with the WFC3 pipeline and the
IRAF\footnote{IRAF is distributed by the National Optical Astronomy
Observatories, which are operated by the Association of Universities
for Research in Astronomy, Inc., under cooperative agreement with the
National Science Foundation.} MULTIDRIZZLE task.  We note that no
narrow-band filter sampling the continuum emission close to the
H$\alpha$ or Pa$\beta$ line was used during the observations. Thus we
use the broad-band {\em V} and {\em I} filters to estimate continuum
in {\em H}$\alpha$ and the medium-band {\em J}-continuum filters for
{\em Pa}$\beta$.

\subsection{H$\alpha$ and Pa$\beta$ Emission}

Radiation emitted at shorter wavelengths is absorbed by dust more
effectively than by photons of longer wavelengths. Given that the {\em
H}$\alpha$ line (restframe wavelength 6563~\AA) is emitted at a shorter
wavelength than {\em Pa}$\beta$ (restframe wavelength 12802~\AA), the
interstellar dust along the line of sight decreases the {\em
H}$\alpha$ flux detected by the observer more than the {\em Pa}$\beta$
flux emitted by the same source. Therefore, the observed {\em
H}$\alpha$/Pa$\beta$ flux ratio is smaller than the theoretical one
computed for the same conditions of electron density ($N_e$) and
temperature ($T_e$) of the source in the absence of dust.

Equation (1) (taken from Calzetti et al.\ 1996) illustrates the
relation between the color excess, $E(B-V)_{\rm g}$ of the
interstellar gas, and the observed and theoretical $H\alpha$/{\em
Pa}$\beta$ flux ratios of this same gas ($R_{\rm obs}$ and $R_{\rm
int}$, respectively) under the assumption of a specific extinction
law (represented by $\kappa(\lambda)$=$A(\lambda)/E(B-V)$):

\begin{equation}
  E(B-V)_g=\frac{-\log(R_{obs}/R_{int})}{0.4\,[\kappa(\lambda_{H\alpha})-\kappa(\lambda_{Pa\beta})]}\,\,\,\rm{mag}
\end{equation}

In order to apply Equation (1), we need to derive the $H\alpha$ and
{\em Pa}$\beta$ emission fluxes per pixel from the available WFC3
images, and for this purpose we make use of IRAF standard routines. We
first determine an average point-spread function (PSF) in each filter
for 15 -- 20 stars in common to all images.  Since the PSF is largest
in the {\em F139M} image, we degrade all the other images by
 convolving them with a Gaussian function whose dispersion, $\sigma^2$, is the difference 
between the PSF $\sigma^2$ of the {\em F139M} image and that of the 
image in the filter in question. We also re-scale the
optical images to the same pixel scale of the near-infrared ones, and
align all images to the one taken through the {\em F139M} filter. We
calibrate all images in flux by multiplying them by their respective
filter PHOTFLAM value as given in the WFC3 manual (Dressel et al.\
2010). We derive the continuum emission at the $H\alpha$ ({\em
Pa}$\beta$) wavelength by interpolating the flux in the {\em F555W}
and {\em F814W} ({\em F127M} and {\em F139M}) images pixel by pixel
with a simple first-order polynomial.  Such an interpolation allows us
to take into account the slope of the continuum emission of the stars
and to better remove the stars from the final, pure line-emission
images.  

The continuum emission derived by interpolating the flux between {\em F555W}
and {\em F814W} ({\em F127M} and {\em F139M}) is subtracted from the {\em
F656N} ({\em F128N)} image, and the output is multiplied by the width of the
narrow-band filter.  In this way we are able to construct the images of the
pure {\em H}$\alpha$ and {\em Pa}$\beta$ emission as well as their observed
$R_{\rm obs}$ flux ratio. We derive $R_{\rm int} = 17.546$ from Osterbrock
(1989) under the assumption that $T_e = 10\,000$~K and $N_e = 100$~cm$^{-3}$.
We adopt a normal extinction law with a ratio of total to selective extinction
of $R_V = 3.1$, and derive $\kappa(\lambda_{H{\alpha}})=2.355$ and
$\kappa(\lambda_{ Pa{\beta}})=0.7644$ from Fitzpatrick's (1999) extinction law.
Finally, we apply Equation (1) to the image of the observed $R_{\rm obs}$ flux
ratio to construct the pixel-to-pixel map of $E(B-V)_{\rm g}$ shown in Figure
1. In panel (a) of Figure 2 we show the histogram of the $E(B-V)_{\rm g}$ per
pixel, normalized by the total number of pixels in the $R_{\rm obs}$ image.
The total line-of-sight color excess is always larger than the foreground
reddening of $E(B-V) = 1.1$ mag (Pandey et al.\ 2000). More than 90\% of the
pixels have $E(B-V)_{\rm g}$ between 1.6 and 2.2~mag.

\subsection{Uncertainty of $E(B-V)_{\rm g}$}

In order to estimate the accuracy of our derived $E(B-V)_{\rm g}$
values, we assume that the electron counts in the images follow a
Poisson distribution.  Therefore we start with a Poisson error on the
electron counts per pixel in the input images, and propagate these
errors through the entire computation of the continua, their
subtraction from the {\em F656N} and {\em F128N} images, and the
computation of the $E(B-V)_{\rm g}$ image according to Equation (1).

We present the histogram of the uncertainties $\sigma_{E(B-V)_{\rm
g}}$ per pixel in panel (b) of Figure 2, normalized by the total number
of pixels in the $R_{\rm obs}$ image.  The histogram peaks at
$\sigma_{E(B-V)_{\rm g}} = 0.1$ mag, and the percentage of pixels with
$\sigma_{E(B-V)_{\rm g}}
> 0.2$ mag is less than 10\%. In panel (c) of Figure 2 we plot the
average $\sigma_{E(B-V)_{\rm g}}$ per pixel in bins of $E(B-V)_{\rm
g}$ per pixel. As can be seen, $\sigma_{E(B-V)_{\rm g}}$ gradually
decreases with increasing $E(B-V)_{\rm g}$ from 0.4 mag at
$E(B-V)_{\rm g} = 1.2$ mag to 0.1 mag at $E(B-V)_{\rm g} = 2.4$ mag.

In addition to photon noise, there are also some systematic errors
because of the following effects:  
\par\noindent (1) Broad-band and medium-band filters are contaminated
by emission lines.  The Pa$\beta$ emission falls in the {\em F127M}
filter. Assuming that the flux density of the continuum is the same in
both the {\em F127M} and {\em F128N} filters, we can use the flux
ratio {\em F127M/F128N} (where both filters were multiplied by their
respective PHOTFLAM and band width) to derive the contribution of the
Pa$\beta$ line to the flux in {\em F127M}. We find this contribution
to be about 13\%.  No hydrogen lines are found in the wavelength range
of the {\em F139M} filter, only a few faint He emission lines, which
should not significantly contribute to the flux in this filter. The
{\em F555W} filter includes four strong emission lines: the H$\beta$
and the [O\,{\sc iii}]$\lambda=4959-5007$~\AA\ line, where the filter
response is $\sim27$\% and the H$\alpha$ line at a response of
$\sim3$\%. The {\em F814W} filter contains the [S\,{\sc iii}] and
[Ar\,{\sc iii}] emission lines at wavelengths where its response is
lower than 10\%.  Unfortunately, we cannot estimate the contribution
of these lines to the flux detected in each filter, because the only
available spectrum taken of NGC\,3603 in the range of 3000 --
10\,400~\AA\/ by Garcia-Rojas et al.\ (2006) was acquired at a
position outside the WFC3 field of view.  If we assume that the line
contamination is about 10\% in {\em F555W} and {\em F814W} as it is
for {\em F127M}, the color excess would then increase by 0.05 mag on
average.  
\par\noindent (2) The adopted extinction law also contributes to the
systematic uncertainties. If we replace Fitzpatrick's (1999)
extinction law with that of Cardelli et al.\ (1989), we obtain an
$E(B-V)_{\rm g}$ systematically smaller by 0.1 mag ($R_V=3.1$). 
 When we vary $R_V$ by $\pm$0.5 in Cardelli et al.'s (1989) extinction law, 
$E(B-V)_{\rm g}$ decreases by $\sim$0.3\,mag ($R_V=3.6$) or
 increases by $\sim$0.2\,mag ($R_V=2.6$).
\par\noindent (3) The assumed electron temperature and density of the
gas may be affected by systematic errors.  G\'arcia-Rojas et al.\
(2006) and Lebouteiller et al.\ (2008) obtained $T_e = 10\,000$~K and
$N_e = 1000$~cm$^{-3}$ for the NGC\,3603 giant H\,{\sc ii} region.
Since this electron density is not available in Osterbrock (1989), we
calculated $R_{\rm int}$ for $T_e = 10\,000$~K and $N_e =
10\,000$~cm$^{-3}$ and derived a new pixel-to-pixel map of
$E(B-V)_{\rm g}$. A factor of 100 difference in $N_e$ results in an
average difference of 0.0025 mag in $E(B-V)_{\rm g}$, which is well
within the errors in $E(B-V)_{\rm g}$ due to photon noise.

\section{A Two-Dimensional Map of the Color Excess}

\subsection{Global Properties of $E(B-V)_{\rm g}$}

The pixel-to-pixel map of $E(B-V)_{\rm g}$ obtained for $T_e =
10\,000$~K and $N_e = 100$~cm$^{-3}$ is presented in Figure 1. The
resulting $E(B-V)_{\rm g}$ is integrated along the different lines of
sight towards NGC\,3603 and includes the Galactic foreground
reddening, which amounts to $E(B-V) = 1.1$~mag according to Pandey et
al.\ (2000).  The HD\,97950 cluster core ($r < 0.52$~pc) where dozens
of OB stars reside is masked out in order to avoid saturated bright
stars in the {\em F555W} and {\em F814W} filters.  Since the PSFs of
the different filters may be slightly different even after the
convolution procedure to obtain the same PSF described in Section 2.1,
the subtraction of the continua from the $H\alpha$ and {\em Pa}$\beta$
images can produce negative values around the stars in the pure
emission images, and thus negative $E(B-V)_{\rm g}$ in the color
excess map. This is particularly true for saturated stars and their
spikes. Therefore, we smooth the color excess map by replacing the
negative $E(B-V)_{\rm g}$ at the position of a star with  the median
value of $E(B-V)_{\rm g}$ in an annulus around it. 

In Figure 1, $E(B-V)_{\rm g}$ is seen to decrease by 0.2~mag when
going from the eastern or the western edge of the field of view to the
cluster center.  Overall, the gas color excess in the east is 0.1~mag
larger than in the west.  However, in the north-south direction
$E(B-V)_{\rm g}$ is systematically larger by 0.4 -- 0.5~mag than in
the east-west direction. In general, the southern region has the
largest color excess as may be expected since here we are moving
towards the densest regions of the giant molecular cloud in the
cluster vicinity (e.g., N\"urnberger et al.\ 2002).  Our findings
support earlier suggestions that the line-of-sight dust density is
uneven across the HD\,97950 cluster surroundings.

As an alternative way of quantifying the differential reddening around
the HD\,97950 cluster, we show in Figure 3 the radial dependence of
$E(B-V)_{\rm g}$ from the cluster core towards the east, west, north,
and south. We define a series of concentric annuli of the same width
(0.2~pc) and with increasing distance from the cluster center, and we
measure the median value and the 16 and 84 percentile values of the
$E(B-V)_{\rm g}$ distribution within each annulus. The solid lines in
Figure 3 trace the median $E(B-V)_{\rm g}$ as a function of
cluster-centric distance, while the long-dashed lines indicate the 84
and 16 percentile values.  Towards the west, the color excess
increases by 0.1~mag from 1~pc to 1.2~pc and stays rather flat at
about 1.55~mag over the distance range 1.2~pc $< r < 3$~pc. In the
eastern direction, $E(B-V)_{\rm g}$ steadily increases from 1.5~mag at
$r = 1$~pc to 1.7~mag at $r = 2$~pc and then remains nearly constant at
1.7~mag for 2~pc $< r < 3$~pc. The most dramatic changes in
$E(B-V)_{\rm g}$ are seen along the northern and southern directions,
where $E(B-V)_{\rm g}$ gets larger by 0.3 to 0.5~mag as the distance
from the cluster center increases.  Specifically, the color excess
increases from about 1.7~mag at $r = 1$~pc to 2.2~mag at a distance of
2~pc.  The pronounced increase towards the south does not come as a
surprise since the overall density of the giant molecular clouds in
NGC\,3603 increases in this direction (see, e.g., Figure 1 in Brandner
et al.\ 1997a).

In the two-dimensional map of the color excess (Figure 1), some
specific sources are associated with a locally higher $E(B-V)_{\rm
g}$. This is the case for two of the ``proplyd''-like
(protoplanetary-disks-like) objects first detected by Brandner et al.\
(2000). Proplyd 1 is located east of the HD\,97950 star cluster, while
Proplyd 3 is in the northwest. Both are tadpole-shaped and
rim-brightened sources with extended tails pointing away from the
cluster core. 

Mid-infrared observations (N\"urnberger \& Stanke 2003) do not reveal
any point-like source that might be associated with the proplyds.
These authors suggest that these objects are small dense clumps of gas
and dust, which are being photoevaporated by the intense ionizing
radiation of the massive stars in the HD\,97950 cluster instead of
being disks around young stars as proposed by Brandner et al.\ (2000).
In particular, Proplyd 3 shows extended faint emission at 11.9~$\mu$m,
which may be caused by carbon-rich dust grains (see Goebel et al.\
1995).  Assuming that the dust properties are the same for these three
proplyds, the higher color excess of Proplyd 3 (0.3 mag redder than
Proplyd 1, see Table 2) would be consistent with a higher dust content
(if both proplyds are at the same distance), and hence with the fact
that Proplyd 3 is the only proplyd detected at 11.9 $\rm \mu m$
(N\"urnberger \& Stanke 2003).  

To the north of the HD\,97950 cluster, we find Sher 25, a blue
supergiant with a circumstellar ring and an hourglass nebula (Brandner
et al.\ 1997a, 1997b).  We have measured the color excess in the
bright rims of the northeastern and southwestern lobes of the hourglass nebula
 (Table 2; see also Figure 1) and found
that the hourglass nebula is more strongly extincted in the northeast (with
$E(B-V)_{\rm g} = 2.0$~mag versus $E(B-V)_{\rm g} = 1.6$~mag in the
southwest).

\subsection{The Shell Structure}

Optical images of NGC\,3603 show an extended cavity around the
HD\,97950 cluster. Clayton (1986) estimated the diameter of this
shell-like structure to be approximately 2 kpc.  Clayton analyzed the
Doppler shifts of the H$\alpha$ emission in the central region of
NGC\,3603, and found a complex velocity structure in the ionized
hydrogen.  He interpreted the main velocity components as indicative
of two expanding shells.  One of these is a faint shell in the
east-west direction with a mean heliocentric velocity of $\langle v
\rangle_{\rm hel} = +120$~km~s$^{-1}$, an expansion velocity of $\sim
30$~km~s$^{-1}$, and a diameter of $\sim 1$~pc. The other one is a
brighter shell in the north-south direction and centered on the
HD\,97950 cluster.  For this second shell-like feature Clayton (1986)
found $\langle v \rangle_{\rm hel} = +50$~km~s$^{-1}$ and $\sim 1.5$
pc as a lower limit for its diameter.  (Note that Clayton assumed that
NGC\,3603 lies at a distance of 6 kpc from us.) From Clayton's Figure
4, we infer an expansion velocity of $\sim 55$~km~s$^{-1}$ for this
second shell, which Clayton identified with the above mentioned
cavity. 

In the $E(B-V)_{\rm g}$ map (Figure 1) we also find a cavity in
correspondence with this expanding gas shell and with a radius of
about 1.0~pc. The median color excess within the cavity is
$E(B-V)_{\rm g} = 1.51 \pm 0.04$~mag. Moreover, we find a shell
structure corresponding to a local enhancement of color excess with a
mean $E(B-V)_{\rm g} = 1.59$~mag about 1.2~pc west of the cluster
center (see Figure 4). This shell structure extends for 1.3~pc along
the north-south direction (Figure 1). Judging from its position, it
seems to lie along the high ionization front of the gas shell
structure found by Clayton (1986) in the north-south direction.
 
Recently, Lebouteiller et al.\ (2007) observed NGC\,3603 with Spitzer
in order to search for PAH (polycyclic aromatic hydrocarbon) and
[S\,{\sc iv}] emission.  They found that within 1.5~pc from the
cluster center, the PAH and [S\,{\sc iv}] emission is suppressed,
possibly because the strong radiation field of the massive stars in
the cluster core (about 60 OB stars, see Melena et al.\ 2008) destroys
dust grains, dissociates complex molecules and pushes the gas out. In
doing so, these stars would thus reduce the color excess within the
area and produce a cavity with lower $E(B-V)_{\rm g}$ in the reddening
map. The western edge of this cavity appears to be delineated by the
shell structure seen in the reddening map.

\subsection{The Molecular Pillars}

In the southern region of NGC 3603 we can see two prominent pillars to
the southwest and southeast of the cluster (Figures 1 and 4).
Following N\"urnberger et al.\ (2002), we refer to them as MM1 and
MM2, respectively.  Figure 4 is a composite color image showing  the
H$\alpha$ emission (in blue), the Pa$\beta$ emission (in green), and
the color excess (in red). The heads of both MM1 and MM2 stand out
because of their strong H$\alpha$ and Pa$\beta$ emission.  They are
gradually being photoevaporated by the cluster's massive OB stars as
is also indicated by the shocked and ionized material in their heads
(N\"urnberger et al.\ 2002, 2003). 

According to Bertoldi (1989) and Bertoldi \& McKee (1990), in clouds
experiencing photoevaporation a velocity gradient will eventually
emerge, leading to a configuration where the head of a pillar moves
more slowly than its tail.  Applying this to the Eagle nebula (M16),
Pound (1998) points out that the measured radial velocity gradients
depend on our viewing angle and provide clues about the
three-dimensional structure of the nebular features with respect to
the ionizing stars.  The tail of a pillar located in front of the
ionizing stars will then have lower radial velocities than the
pillar's head, since the tail is being blown toward the observer
(while the opposite radial velocity gradient would be observed across
a pillar located behind the ionizing OB stars).  Both N\"urnberger et
al.\ (2002) and R\"ollig et al.\ (2010) found that the radial velocity
in the head of MM1 is larger than that of the tail.  Thus in NGC\,3603
we have a configuration where the pillar is located in front of the
stars of the HD\,97950 cluster as seen from our position, with some
tilt with respect to the light of sight.  

The rims of MM1 and MM2 have a high $E(B-V)_{\rm g}$ of $\sim
1.8\,$~mag, implying a large amount of dust. Along the pillar's main
body, though, the color excess is lower than at the rim ($\sim
1.5\,$mag) owing to ``limb brightening'' effects.  We explain this
phenomenon in a cartoon in Figure 5.  The pillar is represented by a
cylinder whose density decreases from the head to the tail (Mackey \&
Lim 2010).  The radiation of the cluster's OB stars ionizes the outer
layer of the pillar (A, B and C in Figure 5), while the interior is
largely shielded and mostly neutral.  Because of the long path
traveled by the light emitted in A and B of the ionized layer (Figure
5), the color excess measured in the pillar rims is relatively high,
yet a lower limit to the true $E(B-V)_{\rm g}$ associated with the
main body of the pillar. The light emitted in C (i.e., the front side
of the pillar in Figure 5) travels a shorter path within the ionized
layer, and, for this reason, the $E(B-V)_{\rm g}$ measured in the
front side of the pillar facing the observer is lower than that
computed at the pillar rims.  

Also in the simulations of Gritschneder et al.\ (2010) the ionizing
radiation emitted by a massive star can enhance the initial turbulent
density distribution of a nearby molecular cloud and naturally lead to
the formation of pillar-like structures within the cloud. The heads of
these pillars may experience gravitational collapse and undergo star
formation. Thompson et al.\ (2002) detected two bright stars in the
head of the pillar column III in M16 using near-infrared observations.

We attempt to estimate the column density of atomic hydrogen ($N_{\rm
H}$) per pixel across the WFC3 field of Figure 1 by using the empirical
relation (Equation 2) of Seward (1999): 
\begin{equation}
N_{\rm H}=1.9\cdot 10^{21}A_V \,\rm atoms~cm^{-2}\,mag^{-1}
\end{equation}

\par\noindent where $A_V$ is computed from the two-dimensional map of
$E(B-V)_{\rm g}$ in Figure 1 assuming $A_{\rm V} = 3.1\cdot
E(B-V)_{\rm g}$. The histogram of the pixel $\log_{10}(N_{\rm H})$
distribution is shown in Figure 6. The majority of the pixels have
$\log_{10}(N_{\rm H})$ in the range of approximately 21.9 to 22.1. The
rims of the pillars in NGC\,3603 are characterized by
$\log_{10}(N_{\rm H}) \simeq 22$. This value is probably only a lower
limit to the true column density of $N_{\rm H}$ within the pillars.
For example, Thompson et al.\ (2002) derived $\log_{10}(N_{\rm H}) =
22.8$ inside the dark region of the pillar column III in M16. 

Since the column density of molecular hydrogen ($N_{\rm H_2}$) is one
half of the $N_{\rm H}$ value (Thompson et al.\ 2002), the pillar rims
in NGC\,3603 likely have  $\log_{10}(N_{\rm H_2}) \simeq 21.7$, which
agrees well with the simulation results of Mackey \& Lim (2010). With
this column density at the rims, the central H$_2$ column density of
the pillar, $\log_{10}(N_{\rm H_2})$, may be as high as about 22.7 to
23 according to Mackey \& Lim (2010), which agrees with the the column
density of MM1 and MM2 ($\log_{10}(N_{\rm H_2})=22.6-23$) as derived
by N$\rm \ddot{u}$rnberger et al.\ (2002).

This value is consistent with the H$_2$ column density of
bright-rimmed clouds (BRCs) undergoing star formation. Urquhart et
al.\ (2009) studied a number of BRCs, which are isolated molecular
clouds located on the edges of evolved H\,{\sc ii} regions. Some BRCs
show evidence of significant interaction between their molecular gas
with the ionizing radiation from their nearby H\,{\sc ii} regions.
These BRCs are thus comparable in their properties to the ionized
heads of molecular clouds. They have been found to host star formation
activity, most likely triggered by the ionizing front coming from
the nearby H\,{\sc ii} regions. The column density of molecular
hydrogen in BRCs with triggered star formation is $20.9 \leq
\log_{10}(N_{\rm H_2}) \leq 22.8$ (Urquhart et al.\ 2009).  

The similarity in $\log_{10}(N_{\rm H_2})$ between star-forming BRCs
and the pillars in NGC\,3603 suggests that the heads of MM1 and MM2
may be sites of star formation.  Indeed Caswell et al.\ (1989) and De
Pree et al.\ (1999) found methanol and water maser sources in the
heads of both MM1 and MM2; these sources are the typical signatures of
newborn stars obscured by dusty molecular envelopes.  The star
formation in the pillars is likely triggered by
photoionization-induced shocks due to the expansion of the H\,{\sc ii}
region in which the HD\,97950 cluster resides.  The strong H$\alpha$
and Pa$\beta$ emission in heads of MM1 and MM2 is another indication
that the OB stars in the HD\,97950 cluster are ionizing the pillars.
The ionization shock fronts were detected at mid-infrared wavelengths
by N\"urnberger \& Stanke (2003).

\section{Summary}

In the present paper we extend previous studies of the differential
reddening around the young star cluster HD\,97950 in the NGC\,3603
giant H\,{\sc ii} region (e.g., Sung \& Bessel 2004).  We derive a
two-dimensional map of the color excess of the gas around the cluster
from {\em HST}/WFC3 images, which allows us to measure the H$\rm
\alpha$/Pa$\rm \beta$ flux ratio and its decrement due to dust
extinction (see Calzetti et al.\ 1996). Our reddening map covers an
area of 4.9~pc $\times$ 4.3~pc with a scale of 0.004~pc per pixel
(assuming that NGC\,3603 is at a distance of 6.9~kpc).  The median
value of the color excess within the central cavity ($r < 1$~pc, where
the HD\,97950 cluster resides) is $E(B-V)_{\rm g}=1.51\pm0.04$ mag.
Going from the cavity to either the northern or the southern edge of
the field of view, $E(B-V)_{\rm g}$ is seen to increase from 1.5~mag
to 2.2~mag at a distance of 2~pc from the cluster. From the cluster
towards either the eastern or western edge of the field of view (at a
distance of 2~pc), $E(B-V)_{\rm g}$ rises from 1.5~mag to 1.6~mag.
The average uncertainty of the derived color excess is about 0.1~mag. 

We find a shell structure 1.2~pc west of the cluster with a mean
$E(B-V)_{\rm g}$ of 1.59~mag, about 0.08~dex higher than in the
central cavity. We interpret this shell structure as the surface at
which the expanding gas shell detected by Clayton (1986) in the
north-south direction interacts with a denser molecular cloud.  

The ionizing radiation emitted by the OB stars in the HD\,97950
cluster is likely to be responsible for the formation of the two
molecular pillars MM1 and MM2 seen 1.2 -- 2.5~pc south-west and
south-east of the cluster, respectively.  We use our reddening map to
estimate the column density of H$_2$ in MM1 and MM2.  We derive
$\log_{10}(N_{\rm H_2})=21.7$ in the pillars' rims and up to
$\log_{10}(N_{\rm H_2})=23$ in the pillars' center, in agreement with
the earlier estimates by N\"urnberger \& Stanke (2003).  Based on the
velocity gradient detected in earlier studies, we argue that the
pillars are closer to us than the ionizing HD\,97950 cluster.  

The strong H$\alpha$ and Pa$\beta$ emission in the heads of the
pillars MM1 and MM2 traces the ionization of the pillar heads by the
OB stars in the HD\,97950 cluster.  The pillar heads appear to be
undergoing star formation as indicated by the presence of methanol and
water maser sources (Caswell et al.\ 1989; De Pree et al.\ 1999).
Such a star formation activity is likely triggered by
photoionization-induced shocks due to the expansion of the H\,{\sc ii}
region surrounding the HD\,97950 cluster.  

The two-dimensional map of $E(B-V)_{\rm g}$ derived in this work 
paves the way to deredden individual stars in the same field of view 
by providing an upper limit to the true stellar reddening. 
Individual reddening corrections in regions
suffering substantial differential reddening are essential for
analyses of stellar photometry, e.g., for constraining the the age
spread in the HD\,97950 cluster and its surroundings.

\acknowledgments The HST/WFC3 multi-wavelength images were processed
by Max Mutchler (STScI) and were kindly provided by Guido De Marchi
and Giacomo Beccari.  We thank Jay Gallagher, Ignacio Ferreras, and
Jorge Garcia-Rojas for helpful discussions.  We also thank the referee 
for useful comments. X.P.\ acknowledges
support in the framework of the Excellence Initiative by the German
Research Foundation (DFG) through the Heidelberg Graduate School of
Fundamental Physics (grant number GSC 129/1). This work was partially 
supported by the Sonderforschungsbereich ``The Milky Way System'' (SFB 881, 
subproject B5) of the DFG.

\clearpage

\onecolumn

\begin{figure}[h]
\includegraphics[width=7cm]{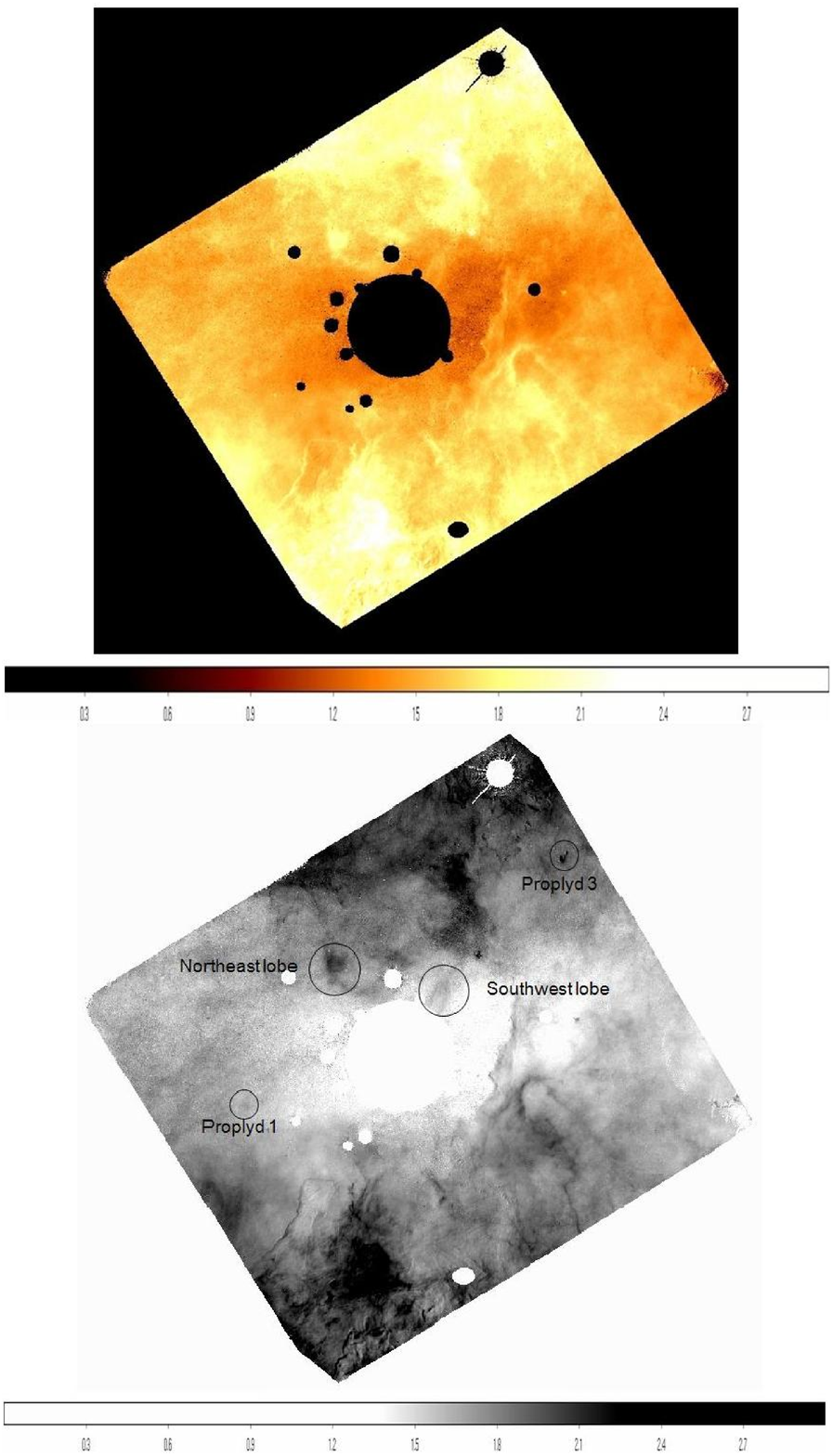}
\caption{Upper panel: The two-dimensional map of the gas color excess,
$E(B-V)_{\rm g}$, around the young cluster HD\,97950 in the
giant H\,{\sc ii} region NGC\,3603.  North is up and east is to the
left. The image covers an area of 4.9~pc $\times 4.3$~pc with a scale
of 0.004~pc per pixel at a distance of 6.9~kpc.  The inner 120~pixels
($r < 0.52$ pc) are masked out due to the luminous OB stars of the
HD\,97950 cluster, which are saturated in the {\em F555W} and {\em
F814W} images obtained with {\em HST}/WFC3. The color excess is
significantly lower along the east-west direction (1.5 -- 1.6~mag)
than in the north-south direction (where it is as high as 2.2~mag on
average).  A shell structure is seen westward of the cluster
associated with a local enhancement of the color excess.  The molecular 
pillars MM1 and MM2 are visible at 1.2 -- 2.5~pc southwest and 
southeast of the HD\,97950 cluster.
 Lower panel: The same map in grey scale. The locations of specific sources 
(Section 3.1) associated with a local $E(B-V)_{\rm g}$ enhancement are indicated.
\label{Fig.1} }
\end{figure}

\begin{figure}[h]
\includegraphics[width=15cm]{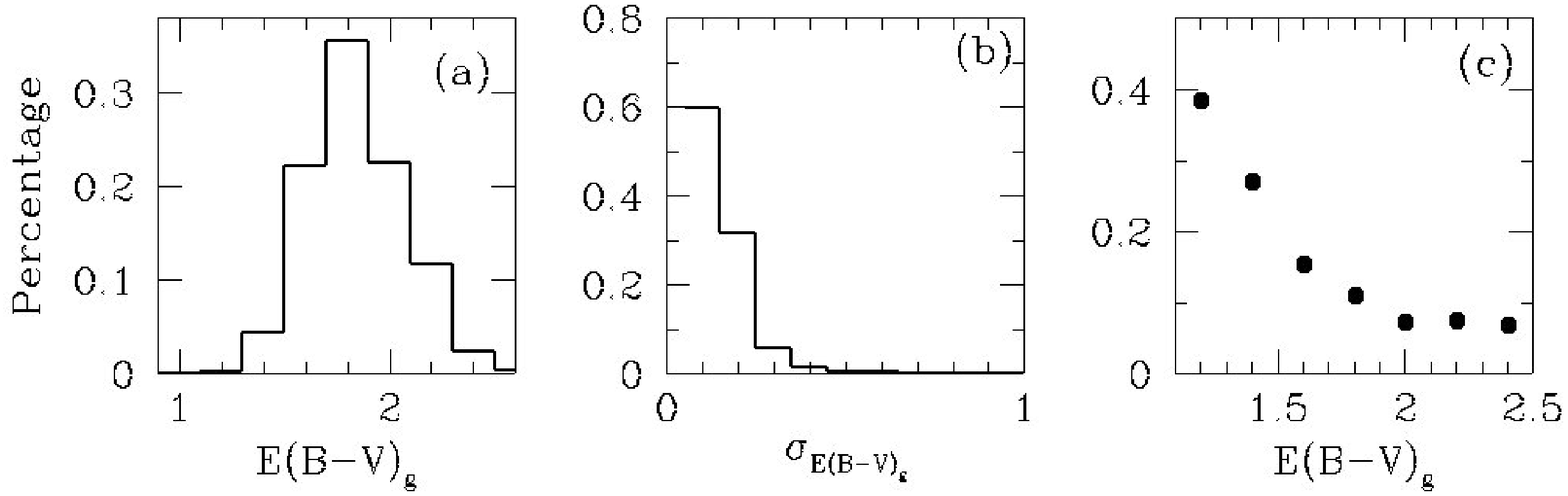}
\caption{Panel a: Histogram of the pixel $E(B-V)_{\rm g}$. Panel b:
Histogram of the uncertainty $\sigma_{E(B-V)_{\rm g}}$ of the pixel
$E(B-V)_{\rm g}$ due to photon noise. Panel c: Dependence of
$\sigma_{E(B-V)_{\rm g}}$ on $E(B-V)_{\rm g}$.
\label{Fig.2}}
\end{figure}

\begin{figure}[h]
\includegraphics[width=12cm]{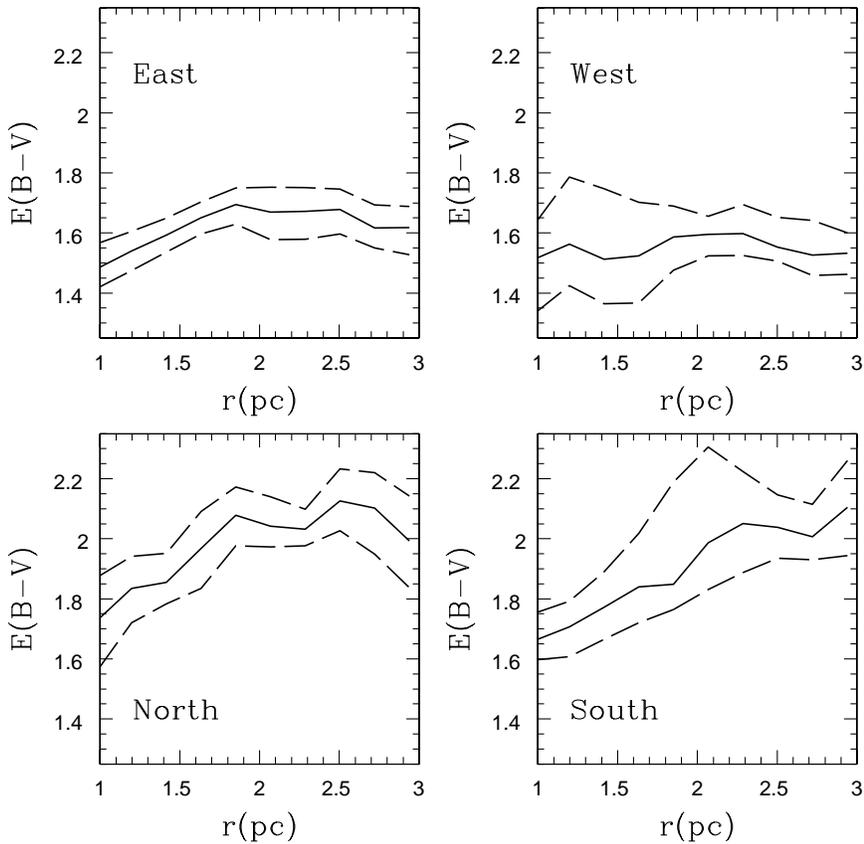}
\caption{The radial dependence of $E(B-V)_{\rm g}$ from the cluster
core towards east, west, north, and south. The solid line is the median
$E(B-V)_{\rm g}$ value, while the upper and lower long-dashed lines
are the 84 and 16 percentile values, respectively, of the $E(B-V)_{\rm
g}$ distribution as a function of distance from the cluster.
\label{Fig.3}}
\end{figure}

\begin{figure}[h]
\includegraphics[width=11cm]{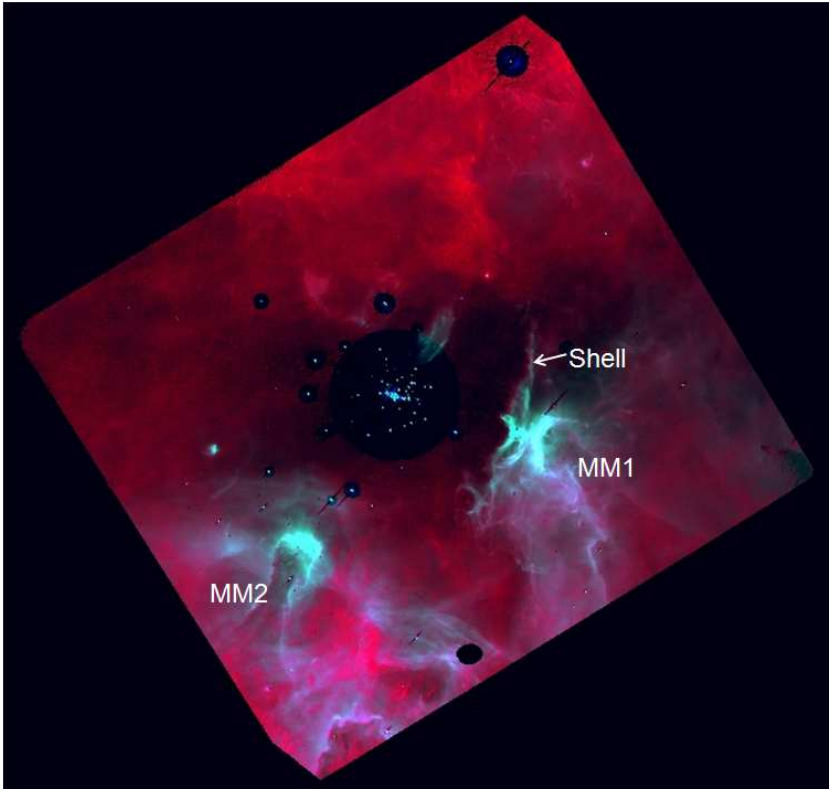}
\caption{Color composite image of NGC\,3603, obtained with the map of
the color excess (red), the {\em Pa}$\beta$ emission image (green), and the
$H\alpha$ emission image (blue).  The positions of the shell and of
the pillars MM1 and MM2 are indicated in the image. 
\label{Fig.4}}
\end{figure}

\begin{figure}[h]
\includegraphics[width=12cm]{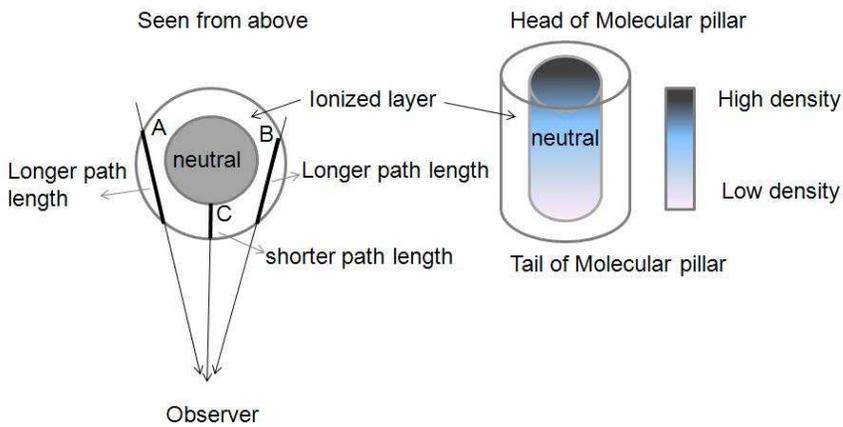}
\caption{A cartoon to illustrate why the observed $E(B-V)_{\rm g}$ in
the rims of the pillar is larger than in the center. The pillar is
represented by a cylinder (neutral in its interior and with a 
density gradient from head to tail) surrounded by a outer layer 
of gas ionized by the OB stars of the HD\,97950 cluster. 
See the text for a more detailed description.
\label{Fig.5}}
\end{figure}

\begin{figure}[h]
\includegraphics[width=10cm]{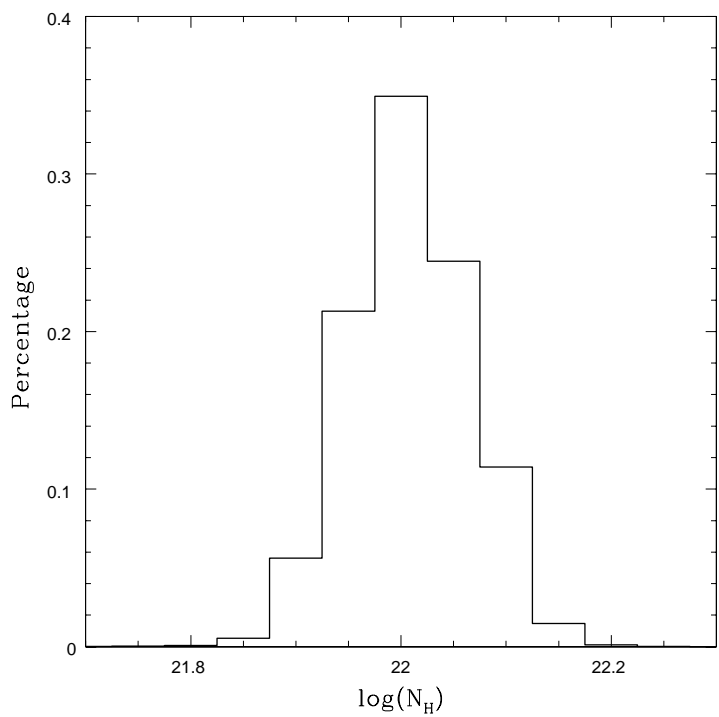}
\caption{Histogram of the pixel column density of atomic hydrogen, 
which is derived from the color excess of each pixel through the 
empirical relation of Seward (1999).
\label{Fig.6}}
\end{figure}

\clearpage

\begin{deluxetable}{rrr}
\tablecolumns{3}
\tablewidth{0pc}
\tablecaption{WFC3 exposure times per filter}
\tablehead{
\colhead{Filter} & \colhead{}   & \colhead{Exposure time}}
\startdata
F555W &  &  1000s \\
F656N &  &  1550s  \\

F814W &  &  990s   \\

F127M &  &  2397.697s \\

F128N &  &  1197.694s \\

F139M &  &  2397.697s \\
\enddata
\end{deluxetable}

\begin{deluxetable}{rccc}
\tablecolumns{4}
\tablewidth{0pc}
\tablecaption{$E(B-V)_{\rm g}$ of specific objects in NGC\,3603}
\tablehead{
\colhead{Object Name} & \colhead{Right Ascension} & \colhead{Declination}  & \colhead{$E(B-V)_{\rm g}$} \\
\colhead{} & \colhead{[~$^{\rm h}$~~$^{\rm m}$~~$^{\rm s}$~]} &\colhead{[~$\degr$~~$^{\rm m~}$~$^{\rm s}$~]} & \colhead{[mag]}}

\startdata
Proplyd 1 & 11 15 13.13 & $-61$ 15 50.0 & 1.6 \\
Proplyd 3 & 11 15 01.41 & $-61$ 14 45.7 & 1.9  \\

Sher 25: northeast lobe & 11 15 09.71 & $-61$ 15 13.2  & 2.0   \\

Sher 25: southwest lobe & 11 15 05.89 & $-61$ 15 25.0 & 1.6 \\
\enddata
\tablecomments{The designations Proplyd 1 and 3 are adopted from 
Brandner et al.\ (2000), while the coordinates of Sher\,25 are  
from Brandner et al.\ (1997a). All coordinates are J2000 coordinates.}
\end{deluxetable}


\begin{thebibliography}{}

\bibitem[Beccari et al.(2010)]{Beccari2010ApJ.720.1108B} Beccari, G., 
et al.\ 2010, \apj, 720, 1108 

\bibitem[Bertoldi(1989)]{1989ApJ...346..735B} Bertoldi, F.\ 1989,
\apj, 346, 735 

\bibitem[Bertoldi \& McKee(1990)]{1990ApJ...354..529B} Bertoldi, F., 
\& McKee, C.~F.\ 1990, \apj, 354, 529 

\bibitem[Brandner et al.(1997a)]{Brandner1997aApJ.475L.45B} Brandner, W., 
Grebel, E.~K., Chu, Y.-H., \& Weis, K.\ 1997a, \apjl, 475, L45 

\bibitem[Brandner et al.(1997b)]{Brandner1997bApJ.489L.153B} Brandner, W., 
Chu, Y.-H., Eisenhauer, F., Grebel, E.~K., \& Points, S.~D.\ 1997b, \apjl, 
489, L153 

\bibitem[Brandner et al.(2000)]{Brandner2000AJ.119.292B} Brandner, W., 
et al.\ 2000, \aj, 119, 292 

\bibitem[Calzetti et al.(1996)]{Calzetti1996ApJ.458.132C} Calzetti, D., 
Kinney, A.~L., \& Storchi-Bergmann, T.\ 1996, \apj, 458, 132 

\bibitem[Cardelli et al.(1989)]{Cardelli1989ApJ.345.245C} Cardelli, J.~A., 
Clayton, G.~C., \& Mathis, J.~S.\ 1989, \apj, 345, 245 

\bibitem[Caswell et al.(1989)]{Caswell1989AuJPh.42.331C} Caswell, J.~L., 
Batchelor, R.~A., Forster, J.~R., \& Wellington, K.~J.\ 1989, Australian 
Journal of Physics, 42, 331 

\bibitem[Clayton(1986)]{Clayton1986MNRAS.219.895C} Clayton, C.~A.\ 1986, 
\mnras, 219, 895 

\bibitem[Clayton(1990)]{Clayton1990MNRAS.246.712C} Clayton, C.~A.\ 1990, 
\mnras, 246, 712 

\bibitem[Crowther et al.(2008)]{Crowther2008ASPC.388.109C} Crowther, 
P.~A., Lennon, D.~J., Walborn, N.~R., \& Smartt, S.~J.\ 2008, 
in Mass Loss from Stars and the Evolution of Stellar Clusters, 
ASP Conf.\ Ser.\ 391, ed.\ K.\ Werner \& T.\ Rauch (San Francisco, CA: 
ASP), 109 

\bibitem[de Pree et al.(1999)]{1999AJ....117.2902D} de Pree, C.~G., 
Nysewander, M.~C., \& Goss, W.~M.\ 1999, \aj, 117, 2902 

\bibitem[Dressel et al.\ (2010)]{Dressel2010} Dressel, L., et al.\ 2010,
Wide Field Camera 3 Instrument Handbook, Version 3.0 (Baltimore:
STScI)\\ http://www.stsci.edu/hst/wfc3/phot\_zp\_lbn

\bibitem[Drissen et al.(1995)]{Drissen1995AJ.110.2235D} Drissen, L., Moffat, 
A.~F.~J., Walborn, N.~R., \& Shara, M.~M.\ 1995, \aj, 110, 2235 

\bibitem[Eisenhauer et al.(1998)]{Eisenhauer1998ApJ.498.278E} Eisenhauer, F., 
Quirrenbach, A., Zinnecker, H., \& Genzel, R.\ 1998, \apj, 498, 278 

\bibitem[Goebel et al.(1995)]{1995ApJ.449.246G} Goebel, J.~H., 
Cheeseman, P., \& Gerbault, F.\ 1995, \apj, 449, 246 

\bibitem[Grebel(2004)]{Grebel2004ASPC.322.101G} Grebel, E.~K.\ 2004, in 
Formation and Evolution of Massive Young Star Clusters, ASP Conf.\ Ser.\ 
322, ed.\ H.J.G.L.M.\ Lamers, L.J.\ Smith, \& A.\ Nota (San Francisco, 
CA: ASP), 101 

\bibitem[Grebel(2005)]{Grebel2005ASSL.327.153G} Grebel, E.~K.\ 2005, in 
The Initial Mass Function 50 Years Later, ASSL Vol.\ 327, 
ed.\ E.\ Corbelli, F.\ Palla, \& H.\ Zinnecker (Dordrecht: Springer), 
153 

\bibitem[Fitzpatrick(1999)]{Fitzpatrick1999PASP.111.63F} Fitzpatrick, 
E.~L.\ 1999, \pasp, 111, 63 

\bibitem[Garc{\'{\i}}a-Rojas et al.(2006)]{Garca-Rojas2006MNRAS.368.253G} 
Garc{\'{\i}}a-Rojas, J., Esteban, C., Peimbert, M., Costado, M.~T., 
Rodr{\'{\i}}guez, M., Peimbert, A., \& Ruiz, M.~T.\ 2006, \mnras, 368, 253 

\bibitem[Gritschneder et al.(2010)]{Gritschneder2010ApJ.723.971G} 
Gritschneder, M., Burkert, A., Naab, T., \& Walch, S.\ 2010, \apj, 723, 971 

\bibitem[Harayama et al.(2008)]{Harayma2008ApJ.675.1319H} Harayama, Y., 
Eisenhauer, F., \& Martins, F.\ 2008, \apj, 675, 1319 

\bibitem[Lebouteiller et al.(2007)]{Lebouteiller2007ApJ.665.390L} 
Lebouteiller, V., Brandl, B., Bernard-Salas, J., Devost, D., 
\& Houck, J.~R.\ 2007, \apj, 665, 390 

\bibitem[Lebouteiller et al.(2008)]{Lebouteiller2008ApJ.680.398L} 
Lebouteiller, V., Bernard-Salas, J., Brandl, B., Whelan, D.~G., Wu, Y., 
Charmandaris, V., Devost, D., \& Houck, J.~R.\ 2008, \apj, 680, 398 

\bibitem[Mackey \& Lim(2010)]{2010MNRAS.403.714M} Mackey, J., \& Lim, 
A.~J.\ 2010, \mnras, 403, 714 

\bibitem[Melena et al.(2008)]{Melena2008AJ.135.878M} Melena, N.~W., Massey, 
P., Morrell, N.~I., \& Zangari, A.~M.\ 2008, \aj, 135, 878 

\bibitem[Melnick et al.(1989)]{Melnick1989A&A.213.89M} Melnick, J., Tapia,
 M., \& Terlevich, R.\ 1989, \aap, 213, 89 

\bibitem[Moffat(1983)]{Moffat1983A&A.124.273M} Moffat, A.~F.~J.\ 1983, 
\aap, 124, 273 

\bibitem[N{\"u}rnberger et al.(2002)]{Nuernberger2002A&A.394.253N} 
N{\"u}rnberger, D.~E.~A., Bronfman, L., Yorke, H.~W., \& Zinnecker, 
H.\ 2002, \aap, 394, 253 

\bibitem[N{\"u}rnberger \& Stanke(2003)]{2003A&A.400.223N} N{\"u}rnberger, 
D.~E.~A., \& Stanke, T.\ 2003, \aap, 400, 223 

\bibitem[Osterbrock (1989)] {Osterbrock1989book} Osterbrock, D.E. 1989, 
Astrophysics of Gaseous Nebulae and Active Galactic Nuclei, 
(Mill Vally, CA: University Science Books) 

\bibitem[Pandey et al.(2000)]{Pandey2000PASJ.52.847P} Pandey, A.~K., Ogura, 
K., \& Sekiguchi, K.\ 2000, \pasj, 52, 847 

\bibitem[Pang et al.(2010)]{Pang2010IAUS.266.24P} Pang, X., Grebel, 
E.~K., \& Altmann, M.\ 2010, in Star clusters: basic galactic building
blocks throughout time and space, IAU Symp.\ 266, ed.\ R.\ de Grijs, 
\& J.\ Lepine (Cambridge: CUP), 24 

\bibitem[Pound(1998)]{1998ApJ...493L.113P} Pound, M.~W.\ 1998, \apjl, 493, 
L113 

\bibitem[Rochau et al.(2010)]{Rochau2010ApJ.716L.90R} Rochau, B., Brandner, 
W., Stolte, A., Gennaro, M., Gouliermis, D., Da Rio, N., Dzyurkevich, N., 
\& Henning, T.\ 2010, \apjl, 716, L90 

\bibitem[Sagar et al.(2001)]{Sagar2001MNRAS.327.23S} Sagar, R., Munari, U., 
\& de Boer, K.~S.\ 2001, \mnras, 327, 23 

\bibitem[Schnurr et al.(2008)]{Schnurr2008MNRAS.389L.38S} Schnurr, O., 
Casoli, J., Chen{\'e}, A.-N., Moffat, A.~F.~J., \& St-Louis, N.\ 2008, 
\mnras, 389, L38 

\bibitem[Seward (1999)] {Seward1999book} Seward, F.D. 1999, in Allen's 
Astrophysical Quantities, ed.\ A.N.\ Cox (New York: Springer), 197

\bibitem[Sung \& Bessell(2004)]{Sung2004AJ.127.1014S} Sung, H., \& 
Bessell, M.~S.\ 2004, \aj, 127, 1014 

\bibitem[Tapia et al.(2001)]{Tapia2001RMxAA.37.39T} Tapia, M., Bohigas, J., 
P{\'e}rez, B., Roth, M., \& Ruiz, M.~T.\ 2001, \rmxaa, 37, 39 

\bibitem[Thompson et al.(2002)]{Thompson2002ApJ.570.749T} Thompson, R.~I., 
Smith, B.~A., \& Hester, J.~J.\ 2002, \apj, 570, 749 

\bibitem[Urquhart et al.(2009)]{Urquhart2009A&A.497.789U} Urquhart, 
J.~S., Morgan, L.~K., \& Thompson, M.~A.\ 2009, \aap, 497, 789 

\end{thebibliography}
\end{document}